\documentclass[%
 reprint,
showpacs,
 amsmath,amssymb,
 aps,
]{revtex4-1}

\usepackage{dcolumn}
\usepackage{bm}

\usepackage[dvipdfmx]{graphicx,color}

\usepackage{ulem} 

\def\<{\langle}
\def\>{\rangle}
\def\({\left(}
\def\){\right)}
\def\[{\left[}
\def\]{\right]}
\def\up{\uparrow}
\def\dn{\downarrow}

\def\e{\mathrm{e}}
\def\i{\mathrm{i}}
\def\d{\mathrm{d}}
\def\der{\partial}
\def\r{\bm{r}}
\def\k{\bm{k}}
\def\Q{\bm{Q}}

\def\Re {{\mathrm{Re}} \,}
\def\Im {{\mathrm{Im}} \,}

\begin{document}


\title{Memory function approach to in-plane anisotropic resistivity in the antiferromagnetic phase of iron arsenide superconductors} 

\author{Koudai Sugimoto$^1$}
\email{koudai@yukawa.kyoto-u.ac.jp}
\author{Peter Prelov\v{s}ek$^{2,3}$}
\author{Eiji Kaneshita$^4$}
\author{Takami Tohyama$^{1,5}$}%
\affiliation{%
$^1$Yukawa Institute for Theoretical Physics, Kyoto University, Kyoto 606-8502, Japan\\
$^2$J. Stefan Institute, SI-1000 Ljubljana, Slovenia\\
$^3$Faculty of Mathematics and Physics, University of Ljubljana, SI-1000 Ljubljana, Slovenia\\
$^4$National Institute of Technology, Sendai College, Sendai 989-3128, Japan\\
$^5$Department of Applied Physics, Tokyo University of Science, Tokyo 125-8585, Japan
}%

\date{\today}

\begin{abstract}
We theoretically examine anisotropy of in-plane resistivity in the striped antiferromagnetic phase of an iron arsenide superconductor by applying a memory function approach to the ordered phase with isotropic nonmagnetic impurity. We find that the anisotropy of the scattering rate is independent of carrier density when the topology of the Fermi surface is changed after the introduction of holes.
On the other hand, the anisotropy of the Drude weight monotonically decreases reflecting the distortion of the Dirac Fermi surface and eventually leads to the reverse of anisotropy of resistivity, being consistent with experiment. 
The origin of the anisotropy is thus attributed to the interplay of impurity scattering and anisotropic electronic states.​
\end{abstract}



\maketitle



\section{Introduction}

Anisotropic electronic states in iron-based superconductors have intensely been studied in connection with the mechanism of high-temperature superconductivity. 
The anisotropy occurs in the FeAs plane through the breakdown of four-fold symmetry in magnetic~\cite{Kasahara12Nature}, electric~\cite{Chu10Science,Tanatar10PRB, Ishida11PRB, Kuo11PRB, Ying11PRL, Ishida13PRL}, and electronic~\cite{Yi11PNAS,Dusza11EPL, Nakajima11PNAS, Nakajima12PRL, Chuang10Science, Allan13NP, Fu12PRL} properties, resulting in a nematic state with two-fold symmetry distinguishing two Fe-Fe directions.
Such a nematic state emerges above structural transition temperature $T_\mathrm{S}$ and antiferromagnetic (AFM) transition temperatures $T_\mathrm{N}$.
Possible origins of the anisotropy above the transition temperatures have been suggested based on spin fluctuation~\cite{Fernandes11PRL} and orbital ordering~\cite{Chen10PRB,Inoue12PRB} that may contribute to the pairing of electrons. 

Since the AFM phase has electronic states different from that of the paramagnetic (PM) phase, the origin might be distinguished from those in the PM phase.
Resistivity measurements for detwinned samples of Co-doped BaFe$_2$As$_2$~\cite{Chu10Science, Ishida13PRL} have clearly shown that in-plane anisotropy is strongly enhanced below $T_\mathrm{N}$.
Therefore, it is important to understand the origin of the enhanced anisotropy in the AFM phase.
In the AFM phase of parent compound BaFe$_2$As$_2$, the anisotropy almost disappears with annealing the sample~\cite{Ishida11PRB}, while it remains finite in Co-doped samples~\cite{Ishida13PRL}.
This result suggests that the anisotropy of resistivity should be induced by impurity scattering of Co atoms substituted for Fe atoms.
It is important to make clear whether the effect of impurity is crucial in accounting for the anisotropy~\cite{Gastiasoro14PRB} or the Drude weight defined without impurity is enough for this purpose~\cite{Valenzuela10PRL, Zhang11PRB, Liang12PRL}.
In hole-doped compounds (Ba$_{1-x}$K$_{x}$)Fe$_2$As$_2$, the anisotropy disappears~\cite{Ishida13JACS} and reverses~\cite{Blomberg13NC}.
This reversing of the anisotropy has been proposed theoretically by taking into account spin fluctuation in the nematic state of PM phase~\cite{Fernandes11PRL,Blomberg13NC}.
However, the theory cannot be applied in the AFM phase.

In this paper, we theoretically examine the resistivity in the AFM phase of iron-based superconductors at zero temperature.
The AFM state is obtained by a mean-field theory of a five-orbital Hubbard model with widely used hopping parameters~\cite{Kuroki08PRL}.
The Fermi surfaces in the AFM state, including Dirac pockets~\cite{Ran09PRB, Morinari10PRL}, are distributed near the $\Gamma$ point in a doping-dependent manner.
The anisotropy of resistivity obtained by the memory function approach~\cite{Gotze72PRB} taking into account impurity scattering is consistent with the experiments near the undoped region, where the resistivity in the AFM-ordered direction is smaller than that in the ferromagnetically ordered direction.
We find that the anisotropy of the scattering rate is independent of hole concentration, once two electron pockets above and below a $\Gamma$-point hole pocket disappear.
On the other hand, the anisotropy of the Drude weight monotonically decreases reflecting the distortion of the Dirac Fermi surface and eventually leads to the reverse of anisotropy of resistivity.
The reverse has been observed in experiments~\cite{Blomberg13NC}.
These results indicate that the origin of the anisotropy is attributed to the interplay of impurity scattering and anisotropic electronic states, both of which are strongly influenced by the topology of the Fermi surface.

This paper is organized as follows.
In Sec. II  the five-orbital Hubbard model describing iron pnictides is introduced together with the mean-field approximation for the AFM state.
In Sec. III, we introduce the expression of the memory function for multi-orbital systems with nonmagnetic impurity and the relation between the memory function and resistivity.
In Sec. IV, the doping dependence of anisotropy of resistivity together with the Drude weight and scattering rate is calculated for the AFM phase of iron pnictides.
The same method is applied to the paramagnetic state but with orbital ordering in Sec. V.
Conclusions are given in Sec. VI.


\section{Model Hamiltonian}

We start with a multiband Hubbard Hamiltonian for a $d$-electron system $H_d = H_0 + H_I$.
The noninteracting Hamiltonian $H_0$ is given by
\begin{equation}
 H_0	= \sum_{i,j} \sum_{\sigma, \mu, \nu}  \left\{ t(\bm{\Delta}_{i, j}; \mu, \nu) + \varepsilon_\mu \delta_{\mu,\nu} \right\} c_{i, \mu, \sigma}^\dagger c_{j, \nu, \sigma},
\label{H0}
\end{equation}
where $ c_{i, \mu, \sigma}^\dagger $ creates an electron at site $i$ with orbital $\mu$ and spin $\sigma$.
$\bm{\Delta}_{i, j} \equiv \r_i - \r _j$, where $\r_i$ is a position of site $i$.
$\varepsilon_\mu$ and $t(\bm{\Delta}_{i,j}; \mu, \nu)$ are the on-site energies and hopping integrals, respectively.
The interaction Hamiltonian $H_I$ is expressed as~\cite{Oles83PRB}
\begin{align}
 H_I &= U\sum_{i,\mu} c^\dagger_{i, \mu, \up} c_{i, \mu, \up} c^\dagger_{i, \mu, \dn} c_{i, \mu, \dn} \notag \\
      &+ \( U-2J \) \sum_{i,\mu \not= \nu} c^\dagger_{i, \mu, \uparrow} c_{i, \mu, \uparrow} c^\dagger_{i, \nu, \dn} c_{i, \nu, \dn}
	\notag \\
      &+ \frac{U-3J}{2} \sum_{i,\mu \neq \nu,\sigma} c^\dagger_{i, \mu, \sigma} c_{i, \mu, \sigma} c^\dagger_{i, \nu, \sigma} c_{i, \nu, \sigma}
    \notag \\
      &- J  \sum_{i,\mu \neq \nu} \left( c^\dagger_{i, \mu, \up} c_{i, \mu, \dn} c^\dagger_{i, \nu, \dn} c_{i, \nu, \up}
- c^\dagger_{i, \mu, \up} c_{i, \nu, \up} c^\dagger_{i, \mu, \dn} c_{i, \nu, \dn} \right),
\end{align}
where $U$ is the intra-orbital Coulomb interaction, and $J$ is the Hund's coupling.
Here, we assume that the pair hopping is equal to $J$.

We construct the mean-field Hamiltonian $H_d^{\rm MF}$ from $H_d$ and self-consistently solve mean-field equations containing the order parameter defined by $\langle n_{\bm{Q}, \mu, \nu, \sigma} \rangle = N^{-1}\sum_{\bm{k}} \langle c^\dagger_{\bm{k}, \mu, \sigma} c_{\bm{k} + \bm{Q}, \nu, \sigma} \rangle$ with the AFM ordering vector $\bm{Q}$, where $N$ is the number of the lattice points, $ c_{\bm{k}, \mu, \sigma}^\dagger = \frac{1}{\sqrt{N}} \sum_i c_{i, \mu, \sigma}^\dagger \e^{\i \k \cdot \r_i}$, and the average $\langle \cdots \rangle$ is taken at zero temperature.
For AFM state, we can rewrite the sum of the wave vectors as $\sum_{\k} \rightarrow  \sum_{\k_0} \sum_{m = 0, 1}$ and $\k \rightarrow \k_0 + m\bm{Q}$, where the sum of $\k_0$ is over the magnetically reduced Brillouin zone.
We introduce band-quasiparticle operators $\gamma_{\k_0, \epsilon, \sigma}^\dagger$ at band $\epsilon$, which create one quasiparticle state with energy $E_{\k_0, \epsilon, \sigma}$. 
The quasiparticle operators satisfy $H_d^{\rm{MF}} = \sum_{\k_0, \sigma} \sum_{\epsilon} E_{\k_0, \epsilon, \sigma} \gamma^\dagger_{\bm{k}_0, \epsilon, \sigma} \gamma_{\bm{k}_0, \epsilon, \sigma} $ and $c_{\k_0 + m \bm{Q}, \mu, \sigma} = \sum_{\mu} \psi_{\mu,m; \epsilon} (\k_0, \sigma)  \gamma_{\bm{k}_0, \epsilon, \sigma} $.


\section{Multi-orbital Memory Function}

The external field is introduced as Peierls phase in creation and annihilation operators so that
\begin{equation}
 c_{i, \mu, \sigma}^\dagger \rightarrow \e^{-\i \frac{e}{c\hbar} \bm{A} \cdot \r_i} c_{i, \mu, \sigma}^\dagger , \quad c_{j, \nu, \sigma} \rightarrow \e^{\i \frac{e}{c\hbar} \bm{A} \cdot \r_j} c_{j, \nu, \sigma},
\end{equation}
where $\bm{A}$ is a vector potential.
The Hamiltonian is expanded in terms of $\bm{A}$ as
\begin{equation}
 H_d(\bm{A}) = H_d(0) - \frac{1}{c} \bm{j} \cdot \bm{A} - \frac{1}{2c^2} \bm{A} \cdot \tau \bm{A} + O(A^3),
\end{equation}
where $\bm{j}$ is a current operator and $\tau$ is a stress tensor.
In the case of the mean-field Hamiltonian, i.e., $H_d(0)=H_d^\mathrm{MF}$, the current operator is given by
\begin{equation}
 \bm{j}
	= -c \left. \frac{\der H_d}{\der \bm{A}} \right|_{\bm{A} = 0}
	= \sum_{\k_0, \sigma} \sum_{\epsilon, \epsilon'} \bm{J}_{\epsilon, \epsilon'} (\k_0, \sigma)
		\gamma^\dagger_{\k_0, \epsilon, \sigma}  \gamma_{\k_0, \epsilon', \sigma}
\end{equation}
with
\begin{multline}
 \bm{J}_{\epsilon, \epsilon'} (\k_0, \sigma)
	= \frac{\i}{N} \frac{e}{\hbar} \sum_m  \sum_{i, j}  \sum_{\mu,\nu}
		\bm{\Delta}_{i, j} t(\bm{\Delta}_{i, j}; \mu,\nu) \\
		\times\e^{ - \i (\k_0 + m \bm{Q}) \cdot \bm{\Delta}_{i, j}}
			\psi^*_{\mu, m ; \epsilon} (\k_0, \sigma) \psi_{\nu, m ; \epsilon'} (\k_0, \sigma),
\end{multline}
and the stress tensor is given by
\begin{equation}
 {\rm \tau}_{\alpha\beta}
	= \sum_{\k_0, \sigma} \sum_{\epsilon, \epsilon'} {\rm \tau}_{\epsilon, \epsilon'}^{(\alpha,\beta)} (\k_0, \sigma)
		\gamma^\dagger_{\k_0, \epsilon, \sigma}  \gamma_{\k_0, \epsilon', \sigma}
\label{eq:stress_tensor}
\end{equation}
with
\begin{multline}
{\rm \tau}_{\epsilon, \epsilon'}^{(\alpha,\beta)} (\k_0, \sigma)
	= \frac{1}{N} \(\frac{e}{\hbar}\)^2 \sum_m  \sum_{i,j}  \sum_{\mu,\nu}
		\Delta^{(\alpha)}_{i,j}  \Delta^{(\beta)}_{i,j}   t(\bm{\Delta}_{i,j}; \mu,\nu)  \\
  \times\e^{ - \i (\k_0 + m \bm{Q}) \cdot \bm{\Delta}_{i,j}}
		\psi^*_{\mu, m ; \epsilon} (\k_0, \sigma) \psi_{\nu, m ; \epsilon'} (\k_0, \sigma).
\end{multline}
Here, $\Delta^{(\alpha)}_{i,j}$ is the $\alpha$ component of $\bm{\Delta}_{i,j}$.
The optical conductivity is written as~\cite{Jaklic00AIP}
\begin{equation}
  \sigma_{\alpha\beta} (z) = \frac{\i}{\hbar z N} \left\{ \chi_{\alpha\beta} (z) - \hbar \< \tau_{\alpha\beta} \> \right\},
\end{equation}
where $z = \omega + \i \eta$ with infinitesimal constant $\eta$, and
\begin{equation}
 \chi_{\alpha\beta} (z) = \<\< j^{(\alpha)}; j^{(\beta)} \>\>_z = -\i \int^{\infty}_0 \d t \, \e^{\i z t} \< [j^{(\alpha)} (t), j^{(\beta)}(0)] \>
\end{equation}
is the current-current correlation function, where $j^{(\alpha)} (t)$ is a Heisenberg representation of the $\alpha$ component of the current operator $j^{(\alpha)}$.
If we consider the  Hamiltonian $H_d^{\rm MF}$, the correlation function is written as
\begin{equation}
\chi_{\alpha\alpha}(z) 
	=  \sum_{\k_0, \sigma} \sum_{\epsilon, \epsilon'}
		\frac{f(E_{\k_0, \epsilon, \sigma} ) - f(E_{\k_0, \epsilon', \sigma} )}{(E_{\k_0, \epsilon, \sigma} - E_{\k_0, \epsilon', \sigma}) / \hbar +  z}
		\left| J^{(\alpha)}_{\epsilon, \epsilon'} (\k_0, \sigma) \right|^2,
\label{eq:chi}
\end{equation}
where $f$ is a Fermi distribution function.

In order to take the effect of nonmagnetic impurity into account, we introduce an impurity at site $l$ inducing a local potential $I_{\rm imp}$ that acts equally on all of local orbitals and conserves the spins and the orbitals in scattering processes. The Hamiltonian is written as
\begin{align}
H'
	=& I_{\rm imp} \sum_l \sum_{\sigma} \sum_{\alpha} c_{l, \alpha, \sigma}^\dagger c_{l, \alpha, \sigma} \notag \\
	=& \frac{1}{N} \sum_{\k_0, \k_0', \sigma} \sum_{\epsilon, \epsilon'} \sum_l  I_{\epsilon, \epsilon'}^{l} (\k_0,  \k_0')
	\gamma_{\k_0, \epsilon, \sigma}^\dagger \gamma_{\k_0', \epsilon', \sigma}
\end{align}
with
\begin{align}
I_{\epsilon, \epsilon'}^l (\k_0, \k_0')
	=&  I_{\rm imp}  \sum_{\mu, m, m'}  \e^{- \i (\k_0 - \k_0' + (m - m') \Q) \cdot \r_l} \notag \\
		& \times \psi^*_{\mu, m ; \epsilon} (\k_0, \sigma) \psi_{\mu ,m' ; \epsilon'} (\k_0', \sigma).
\label{eq:I}
\end{align}
The total Hamiltonian of our system is thus $H = H^{\rm total} = H_d^\mathrm{MF}+H'$.

In order to calculate resistivity, we make use of a memory function. The memory function in a system with impurities represents a relaxation due to the impurities and its imaginary part corresponds to the scattering rate of the system~\cite{Gotze72PRB}.
Therefore, the memory function should vanish if there is no impurity.
This condition will be satisfied if the memory function can be expanded in terms of the impurity concentration $c$, starting from the first order of $c$ [$O(c)$]~\cite{Gotze72PRB}.
Since we are interested in resistivity, we have to consider the expansion in the limit of zero frequency, i.e., $\omega\rightarrow 0$. Based on these considerations, we introduce a memory function for multiorbital systems, defined by
\begin{equation}
 M_{\alpha\beta} (z) \equiv - \frac{\i z \, \Im \chi_{\alpha\beta} (z)}{ \i \, \Im \chi_{\alpha\beta} (z) + 2N D_{\alpha\beta} (z)},
\label{eq:memory_function}
\end{equation}
where
\begin{equation}
 D_{\alpha\beta} (z) = \frac{1}{2N} \( \Re \chi_{\alpha\beta} (z) - \hbar \< {\rm \tau}_{\alpha\beta} \> \).
 \label{eq:Drude_weight}
\end{equation}
We note that $D_{\alpha\alpha} (0)$ corresponds to charge stiffness or Drude weight.
The real part of the optical conductivity is
\begin{equation}
 \Re \sigma_{\alpha\beta} (z)
	= \frac{2 D_{\alpha\beta}(z)}{\hbar}
		\frac{\Im M_{\alpha\beta} (z)}{ \left\{ \omega + \Re M_{\alpha\beta} (z) \right\}^2 + \left\{ \Im M_{\alpha\beta} (z) \right\}^2}.
 \label{eq:optical_conductivity}
\end{equation}
We note that Eq.~(\ref{eq:optical_conductivity}) leads to a Drude formula with scattering rate $\Im M_{\alpha\beta}$, if $D_{\alpha\beta} (z)$ is independent of $z$. 

$M_{\alpha\beta} (z)$ given by Eq.~(\ref{eq:memory_function}) is slightly different from that for a single-orbital system: $\i \, \Im \chi_{\alpha\beta}$ in the numerator of Eq.~(\ref{eq:memory_function}) corresponds to $\Re \chi_{\alpha\beta} + \i \, \Im \chi_{\alpha\beta}$ in Eq.~(21) of Ref.~\cite{Gotze72PRB}.
It is important to notice that $\Re \chi_{\alpha\beta}$ remains finite even if $c=0$.
This is due to the presence of interband transition in multiorbital systems.
Therefore, removing $\Re \chi_{\alpha\beta}$ in our definition is necessary for satisfying the condition that the memory function should be expanded starting from $O(c)$.

According to Appendix \ref{appendix:jEPH88yRF5Mxx1pH},
\begin{equation}
 z \, \Im \chi_{\alpha \alpha} (z) =
- \frac{1}{\hbar^2} \Im \[ \frac{ \<\< A_\alpha ; A_\alpha \>\>_{z} - \<\< A_\alpha ; A_\alpha \>\>_{0} }{z} \]
\label{eq:z_Im_chi}
\end{equation}
with $A_\alpha = [ j^{(\alpha)}, H^{\rm total}]$. Regarding $H'$ as a perturbation and using Born approximation, we find that the contribution from $[j^{(\alpha)}, H_d^\mathrm{MF}]$ vanishes in the limit of $z\rightarrow 0$.
This guarantees that, in the limit of $z\rightarrow 0$, $M_{\alpha \alpha} (z)$ given by Eq.~({\ref{eq:memory_function}}) can be expanded starting from $O(c)$. In the dilute limit of $c$, i.e., $O(c)$, we obtain
\begin{align}
M_{\alpha \alpha} (0)
	&= -\frac{\i}{2ND_{\alpha\alpha} (0)} \lim_{z \to 0}z \, \Im \chi_{\alpha \alpha} (z) + O(c^2) \notag \\
	&\simeq \sum_{\k_0} M_{\alpha \alpha} (\k_0, 0),
\label{eq:ImM}
\end{align}
and according to Appendix  \ref{appendix:jEPH88yRF5Mxx1pH}, 
\begin{multline}
\Im M_{\alpha \alpha} (\k_0, 0)
	= \frac{\pi c}{2D_{\alpha\alpha}(0)} \sum_{\k_0'} \sum_{\epsilon, \epsilon'} \sum_{\sigma}
		\left| A^{(\alpha), l=0}_{\epsilon, \epsilon'} (\k_0, \k_0', \sigma) \right|^2\\
	\times \delta (E_{\k_0, \epsilon, \sigma} - E_F) \delta (E_{\k_0, \epsilon, \sigma} - E_{\k_0', \epsilon', \sigma}).
\end{multline}

With use of Eq.~(\ref{eq:ImM}), where $\Re M_{\alpha\alpha}(0) = 0$, the resistivity is written as~\cite{Gotze72PRB}
\begin{equation}
\rho_{\alpha} = \frac{1}{\Re \sigma_{\alpha\alpha} (0)} = \frac{\hbar \, \Im M_{\alpha \alpha} (0)}{2 D_{\alpha\alpha} (0) }.
\end{equation}
We evaluate the Drude weight~(\ref{eq:Drude_weight}) from Eq.~(\ref{eq:stress_tensor}) and~(\ref{eq:chi}).


\section{Anisotropy of resistivity in Antiferromagnetic phase}

The on-site energies and the hopping integrals are taken from Kuroki {\it et al.}~\cite{Kuroki08PRL}.
We set $U=1.2$~eV and $J=0.22$~eV to yield a magnetic moment $m=0.8 \mu_B$ ($\mu_B$ is the Bohr magneton) at $n = 6.0$, where $n$ is the electron density and $m = \sum_\mu \langle n_{\bm{Q}, \mu, \mu, \up} - n_{\bm{Q}, \mu, \mu, \dn} \rangle \mu_B$.
This value is chosen to be close to the measured $m$ of BaFe$_2$As$_2$~\cite{Huang08PRL}.
The calculated value of $m$ linearly decreases with increasing $n$~\cite{Morinari10PRL}.
This is because the nesting condition between the Fermi surfaces at the $Y$ and $M$ points in the PM phase becomes worse.
This will lead to an incommensurate AFM state and thus our assumption of the commensurate AFM state may fail.
Actually, $m$ at the $(\pi,0)$-ordered state in our calculation shows a discontinuous change when $n$ exceeds 6.03.

\begin{figure}
\includegraphics[width=8.0cm]{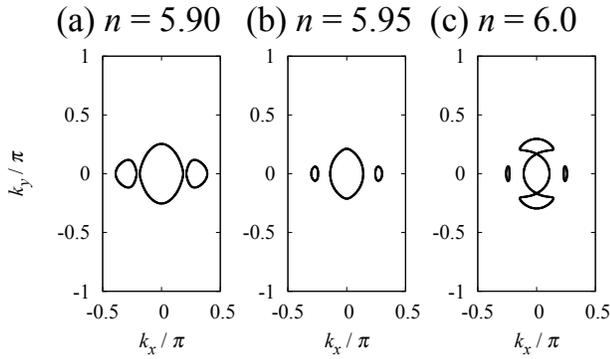}
\caption{
The Fermi surfaces at (a) $n = 5.90$, (b) $n = 5.95$, and (c) $n = 6.0$ in the stripe-type AFM Brillouin zone.
}
\label{fig1}
\end{figure}

The Fermi surfaces at $n =$ 5.90, 5.95 and 6.0 are shown in Figs.~\ref{fig1}(a), \ref{fig1}(b), and \ref{fig1}(c), respectively.
At $n = 6.0$, there are five pockets; a hole pocket at the $\Gamma$ point, two pockets  (Dirac pockets) coming from the Dirac-type linear dispersions on the left and right side of the hole pocket, and two electron pockets above and below the hole pocket.
At $n= 5.98$, the Dirac points meet the Fermi level~\cite{Morinari10PRL}.
At $n = 5.95$, the electron pockets disappear, but the Dirac pockets remain [Fig.~\ref{fig1}(b)]. With further reducing $n$, i.e., increasing hole concentration, the Dirac pockets grow as shown in Fig.~\ref{fig1}(a), since the Dirac points are above the Fermi level.

\begin{figure}
\includegraphics[width=8.0cm]{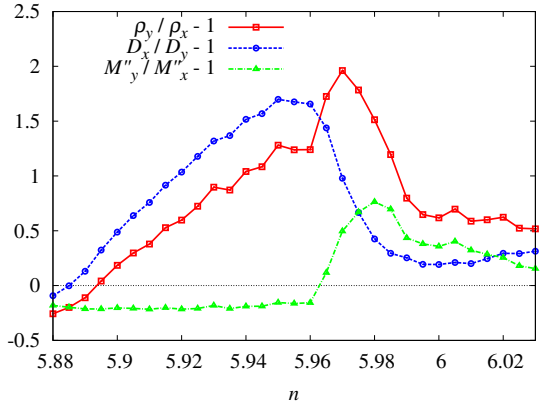}
\caption{
(Color online)
The electron density $n$ dependence of anisotropy of resistivity $\rho_y / \rho_x - 1$ (red squares), Drude weight $D_x / D_y - 1$ (blue circles), and memory function $M''_y / M''_x - 1$ (green triangles).
Here, $D_{x(y)}=D_{xx(yy)}(0)$ and $M''_{x(y)} = \Im M_{xx(yy)}(0)$.
}
\label{fig2}
\end{figure}

Since Co is substituted for Fe in electron-doped Ba(Fe$_{1-x}$Co$_x$)$_2$As$_2$, Co atoms act as impurities.
In hole-doped Ba$_{1-x}$K$_x$Fe$_2$As$_2$, K does not directly affect Fe sites, but may give an additional potential through electrostatic interaction.
Figure~\ref{fig2} shows the anisotropy of resistivity given by $\rho_y / \rho_x - 1$.
We find that $\rho_y > \rho_x$ near $n=6.0$, which agrees with the experiments in electron-doped BaFe$_{2-x}$Co$_x$As$_2$~\cite{Chu10Science, Ishida13PRL}.
The ratio of anisotropy shows a maximum value of 1.9 at $n=5.97$ and decreases with increasing $n$.
This doping dependence is qualitatively consistent with experimental observations, although the maximum of the ratio appears above $n=6.0$ in the experiments~\cite{Ishida13PRL, Blomberg13NC}.
Here, we emphasize that, around undoping ($n=6.0$) region, including impurity scattering makes the in-plane anisotropy of resistivity (red squares) much larger than that of the Drude weight (blue circles).
We thus can say that the anisotropy of the resistivity is enhanced by the effect of impurity scattering described by the memory function. 

Below $n=5.96$, the anisotropy of Drude weight monotonically decreases while the anisotropy of memory function, i.e., the anisotropy of scattering rate, is almost constant with negative sign opposite to that near $n=6.0$. In this region, the two electron pockets at $n=6.0$ disappear, and the Fermi surface has two Dirac pockets and one hole pocket as shown in Fig.~1. This topology of the Fermi surface lasts down to $n=5.88$. Therefore, we may say that the anisotropy of scattering rate depends only on the topology of the Fermi surface, while the anisotropy of Drude weight depends on the details of the Fermi surface such as the curvature of Dirac pockets.

The combination of Drude weight and scattering rate leads to the monotonic decrease of the anisotropy of resistivity below $n=5.96$. The anisotropy is finally reversed in the hole-doped side of $n<5.89$, i.e., $\rho_y < \rho_x$.
The reverse of anisotropy is also consistent with the doping dependence of anisotropy in hole-doped Ba$_{1-x}$K$_x$Fe$_2$As$_2$~\cite{Blomberg13NC}, although the observed carrier concentration where the anisotropy is reversed ($x\sim0.2$) quantitatively disagrees with our results. The agreement with the experiments will be improved by tuning the hopping parameters and by being beyond the dilute limit for large $c$ and the Born approximation.
Furthermore, the commensurability that we have assumed may be getting worse away from $n=6$ as mentioned above, leading to a possible modification of the present doping dependence of anisotropy through the change of Fermi surface topology.
We believe that such factors will give a better agreement with experiments, though there is no specific calculation.
These remain as a future problem.

\begin{figure}
\includegraphics[width=8.0cm]{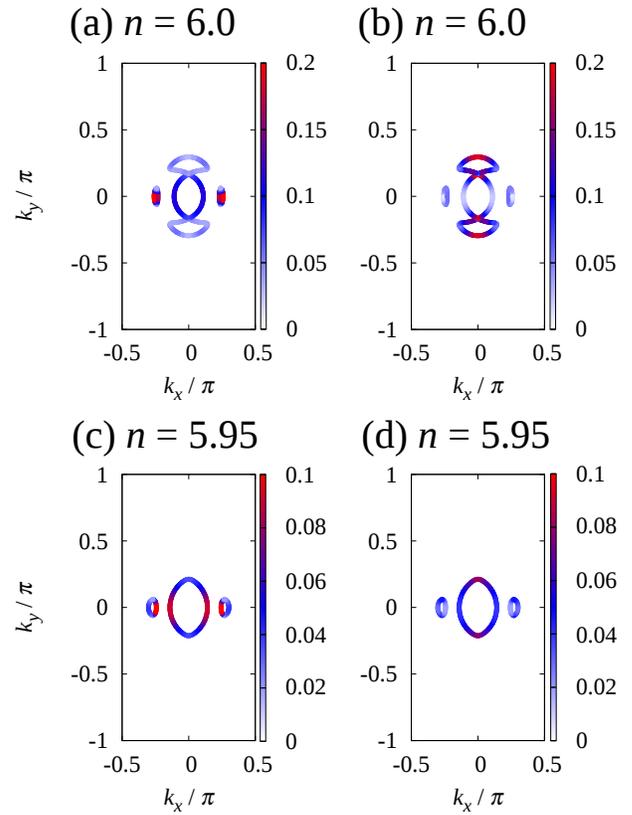}
\caption{
(Color online)
The intensity of $\Im M_{\alpha\alpha}(\k_0, 0)  / (c I_{\rm imp}^2)$.
(a)  $\alpha = x$ and $n = 6.0$.
(b) $\alpha = y$ and $n = 6.0$.
(c) and (d): The same as (a) and (b), respectively, but $n=5.95$.
}
\label{fig3}
\end{figure}

In order to understand the origin of the anisotropy in scattering rate near $n=6.0$ and its doping independent behavior, we plot $\Im M_{\alpha\alpha} (\k_0, 0) / (c I_{\rm imp}^2)$ [see Eq.~(\ref{eq:ImM})] on the Fermi surfaces in Figs.~\ref{fig3}(a), \ref{fig3}(c) and Figs.~\ref{fig3}(b), \ref{fig3}(d) for $\alpha=x$ and $\alpha=y$, respectively.
Figure~\ref{fig3}(a) shows that the scattering rate for $\alpha=x$ mainly comes from the Dirac pockets.
On the other hand, Fig.~\ref{fig3}(b) shows that the electron pockets significantly contribute to the scatting rate for $\alpha=y$.
In other words, electrons are mainly scattered at the Dirac pockets (electron pockets) when current flows along the $x$ ($y$) direction.
At $n=6.0$, we find that the integrated scattering rate over the Fermi surfaces is larger for $\alpha=y$ than for $\alpha=x$.

Since the electron pockets controls the scattering rate for $\alpha=y$, we can expect the reverse of the anisotropy if the electron pockets disappear [see Figs.~\ref {fig1}(a) and \ref{fig1}(b)].
This is actually the case below $n=5.96$ [see Figs.~\ref{fig3}(c) and \ref{fig3}(d)].
These results indicate that both including the impurity scattering and treating the Fermi-surface evolutions are essential for a proper understanding of the origin of the in-plane anisotropy of resistivity.

\section{Anisotropy of resistivity and Energy Separation}

\begin{figure}
\includegraphics[width=8.0cm]{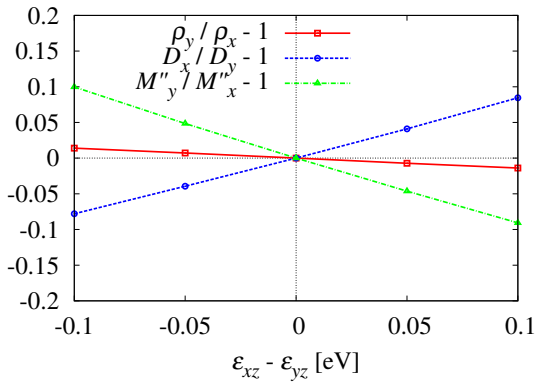}
\caption{
(Color online)
Anisotropy vs energy separation $d_{xz}-d_{yz}$.
Plotted are three quantities indicating anisotropy with respect to the energy separation: the anisotropy of resistivity $\rho_y / \rho_x - 1$ (red squares), Drude weight $D_x / D_y - 1$ (blue circles), and memory function $M''_y / M''_x - 1$ (green triangles).
Here, $D_{x(y)}=D_{xx(yy)}(0)$ and $M''_{x(y)} = \Im M_{xx(yy)}(0)$.
}
\label{fig4}
\end{figure}

In the light of our findings in the AFM state, we discuss the anisotropy in the PM phase.
Above $T_\mathrm{N}$ and $T_\mathrm{S}$, the anisotropy may be related to the nematic state observed experimentally~\cite{Kasahara12Nature}.
A possible mechanism of the nematicity is the orbital ordering and fluctuation.
Angle resolved photo-emission spectroscopy measurements have clearly shown the splitting of $d_{xz}$ and $d_{yz}$ orbitals  ($\varepsilon_{yz}>\varepsilon_{xz}$) in the nematic state~\cite{Yi11PNAS}.
The splitting is an evidence of the orbital ordering.
Based on this observation, we introduce an energy level splitting between $\epsilon_{yz}$ and $\epsilon_{xz}$ in $H_0$~(\ref{H0}) and calculate the Drude weight in the paramagnetic state at $n=6$.
We find that the anisotropy of the Drude weight is opposite to the observed anisotropy of the resistivity, as shown in the $\epsilon_{yz} > \epsilon_{xz}$ region of Fig.~\ref{fig4}.

Then, according to the AFM case, we introduce impurities and apply the memory-function approach, resulting in the anisotropy consistent with experiments, i.e., $\rho_y > \rho_x $ at $\varepsilon_{yz}>\varepsilon_{xz}$ as shown in Fig.~\ref{fig4}.
The interplay of impurity scattering and anisotropic Fermi surfaces due to the orbital ordering has also theoretically been reported by a $T$-matrix formalism~\cite{Inoue12PRB}.
Therefore, we suggest that the orbital ordering is one of the possible origins of the resistivity anisotropy above $T_\mathrm{S}$.


\section{Conclusion}

In summary, we have investigated the in-plane anisotropy of resistivity with multiorbital Hubbard model in the AFM phase.
The resistivity has been obtained from memory function approach.
Both Drude weight and scattering rate tend to make resistivity along the ferromagnetic direction larger than that of AFM direction in parent compound, which is consistent with in-plane anisotropy in experiments.
When holes are introduced, the anisotropy of the scattering rate reverses and becomes independent of carrier concentration, which is related to the disappearance of electron pockets. On the other hand, the Drude weight monotonically decreases as a consequence of deformation of the Dirac pocket with hole doping.
This eventually leads to the reverse of anisotropy of resistivity, which is also consistent with experiments. 
These results indicate that the origin of the anisotropy is attributed to the interplay of impurity scattering and anisotropic electronic states, both of which are strongly influenced by the topology of Fermi surface.
In addition, we have successfully obtained the anisotropy of resistivity consistent with experiments.
These results suggest the importance of impurity scattering in the anisotropy of  resistivity of iron-pnictides.

\begin{acknowledgments}
This work was supported by Grant-in-Aid for Scientific Research from
the Japan Society for the Promotion of Science, MEXT (Grant No. 243649, 22740225, 26400381);
the Global COE Program \lq{}\lq The Next Generation of Physics, Spun from Universality and Emergence\rq{}\rq;
the Strategic Programs for Innovative Research (SPIRE), the Computational Materials Science Initiative (CMSI);
and Yukawa Institutional Program for Quark-Hadron Science.
\end{acknowledgments}

\appendix
\section{Correlation function} \label{appendix:jEPH88yRF5Mxx1pH}

The equation of motion on the Heisenberg-representation operator $O(t)$ is $\frac{\der O(t)}{\der t} = \frac{1}{\i \hbar} [O(t), H^{\rm total}]$.
Integrated by parts, the correlation function of operators $A$ and $B$ becomes
\begin{align}
 z \<\< A; B \>\>_z
 	&= \< [A, B] \> + \frac{1}{\hbar} \<\< [A, H^{\rm total}]; B \>\>_z \notag \\
 	&= \< [A, B] \> - \frac{1}{\hbar} \<\<A; [B, H^{\rm total}] \>\>_z.
\end{align}
When $\alpha = \beta$,
\begin{align}
 z \chi_{\alpha\alpha} (z)
 	&=\frac{1}{\hbar} \<\< A_\alpha; j^{(\alpha)} \>\>_z \notag \\
 	&= \frac{1}{\hbar z} \( \< [A_\alpha, j^{(\alpha)}] \> - \frac{1}{\hbar} \<\< A_{\alpha}; A_\alpha \>\>_z \),
\end{align}
where $A_{\alpha} = [j^{(\alpha)}, H^{\rm total}]$.
Apparently, $\<  [A_\alpha, j^{(\alpha)}]  \> = \frac{1}{\hbar} \<\< A_\alpha; A_\alpha \>\>_{z = 0}$.
Thus, we obtain
\begin{equation}
  z \chi_{\alpha\alpha} (z) = -\frac{1}{\hbar^2 z} \left\{ \<\< A_\alpha; A_\alpha \>\>_{z} - \<\< A_\alpha; A_\alpha \>\>_{z = 0}  \right\}.
\end{equation}

The commutator $A_\alpha$ is explicitly written as
\begin{equation}
 A_{\alpha} =  \sum_{\k_0, \k_0'} \sum_{\sigma} \sum_{\epsilon, \epsilon'} \sum_l A^{(\alpha), l}_{\epsilon, \epsilon'} (\k_0, \k_0', \sigma)
		\gamma_{\epsilon}^\dagger (\k_0, \sigma) \gamma_{\epsilon'} (\k_0', \sigma)
\end{equation}
with
\begin{multline}
 A^{(\alpha), l}_{\epsilon, \epsilon'} (\k_0, \k_0', \sigma)
	= \left\{ E_{\k_0, \epsilon', \sigma} -  E_{\k_0, \epsilon, \sigma}   \right\} J^{(\alpha)}_{\epsilon, \epsilon'} (\k_0, \sigma)  \delta_{\k_0, \k_0'} \\
	+ \frac{1}{N} \left\{  J^{(\alpha)}_{\epsilon, \epsilon} (\k_0, \sigma) I_{\epsilon', \epsilon'}^l (\k_0, \k_0')
		-  I_{\epsilon, \epsilon'}^l (\k_0, \k_0') J^{(\alpha)}_{\epsilon', \epsilon'} (\k_0', \sigma)  \right\}.
\label{eq:1OIPXj0SABfoqA2P}
\end{multline}
Note that we ignore the inter-band matrix elements of the current operator taking the view that the elastic scattering is prohibited in potential impurity scattering.
Its correlation function is
\begin{multline}
   \<\< A^{(\alpha)} ; A^{(\alpha)} \>\>_z 
	=   \sum_{\k_0, \k_0'} \sum_{\sigma} \sum_{\epsilon, \epsilon'}
		\left| \sum_l A^{(\alpha), l}_{\epsilon, \epsilon'} (\k_0, \k_0' , \sigma) \right|^2 \\
		\times \frac{f(E_{\k_0', \epsilon', \sigma}) - f(E_{\k_0, \epsilon, \sigma})}
			{ z - (E_{\k_0', \epsilon', \sigma} - E_{\k_0, \epsilon, \sigma}) / \hbar}
\end{multline}
within Born approximation.

We assume $z \, \Im \chi_{\alpha\alpha} (z) = \Im \left\{ z \chi_{\alpha\alpha} (z) \right\}$ since $z = \omega + \i \eta$, where $\eta$ is an infinitesimal value.
We can calculate
\begin{multline}
 z \, \Im \chi_{\alpha\alpha} (z)
 	= \frac{\pi}{\hbar\omega} \sum_{\k_0, \k_0'} \sum_{\sigma} \sum_{\epsilon, \epsilon'}
 		\left| \sum_l A^{(\alpha), l}_{\epsilon, \epsilon'} \right|^2 \\
 			\times \left\{ f(E_{\k_0', \epsilon', \sigma}) - f(E_{\k_0, \epsilon, \sigma}) \right\}
 				\delta \( \hbar \omega - (E_{\k_0', \epsilon', \sigma} - E_{\k_0, \epsilon, \sigma})  \).
 \label{eq:0K3KGNoOraQo0wTQ}
\end{multline}
In the limit of $\omega \to 0$, the right-hand side of Eq.~({\ref{eq:0K3KGNoOraQo0wTQ}}) becomes
\begin{equation}
 \pi \sum_{\k_0, \k_0'} \sum_{\sigma} \sum_{\epsilon, \epsilon'}
 		\left| \sum_l A^{(\alpha), l}_{\epsilon, \epsilon'} \right|^2
 			\delta \( E_{\k_0', \epsilon', \sigma} - E_F  \)
 			\delta \(E_{\k_0', \epsilon', \sigma} - E_{\k_0, \epsilon, \sigma} \).
 \label{eq:ohRUPa4GDcZsigq6}
\end{equation}
Note that $\frac{\der f(E_{\k_0, \epsilon, \sigma})}{\der E_{\k_0, \epsilon, \sigma}} = -\delta (E_{\k_0, \epsilon, \sigma} - E_F)$ at absolute zero temperature.

The $\delta$-function in Eq.~(\ref{eq:ohRUPa4GDcZsigq6}) implies that the first term in Eq.~(\ref{eq:1OIPXj0SABfoqA2P}) does not contribute to the final result.
In the dilute impurity concentration, we can approximate
\begin{align}
 \left| \sum_l A^{(\alpha), l}_{\epsilon, \epsilon'} \right|^2
 	&= \sum_l \left| A^{(\alpha), l}_{\epsilon, \epsilon'} \right|^2 +  \sum_{l \neq l'} \left| A^{(\alpha), l}_{\epsilon, \epsilon'} A^{(\alpha), l'}_{\epsilon, \epsilon'} \right| \notag \\
 	&\simeq N_i \sum_l \left| A^{(\alpha), l=0}_{\epsilon, \epsilon'} \right|^2 ,
\end{align}
where $N_i$ is the number of the impurity sites.
The second term is neglected because its order is $c^2$, where $c = N_i / N$ is the impurity concentration.

\nocite{*}



\end{document}